\documentstyle[12pt]{article}
\begin{document}
\thispagestyle{empty}
\newcommand{\be}{\begin{equation}}
\newcommand{\ee}{\end{equation}}
\newcommand{\sect}[1]{\setcounter{equation}{0}\section{#1}}
\newcommand{\vs}[1]{\rule[- #1 mm]{0mm}{#1 mm}}
\newcommand{\hs}[1]{\hspace{#1mm}}
\newcommand{\mb}[1]{\hs{5}\mbox{#1}\hs{5}}
\newcommand{\bea}{\begin{eqnarray}}
\newcommand{\eea}{\end{eqnarray}}
\newcommand{\wt}[1]{\widetilde{#1}}
\newcommand{\ux}[1]{\underline{#1}}
\newcommand{\ov}[1]{\overline{#1}}
\newcommand{\sm}[2]{\frac{\mbox{\footnotesize #1}\vs{-2}}
           {\vs{-2}\mbox{\footnotesize #2}}}
\newcommand{\prt}{\partial}
\newcommand{\eps}{\epsilon}\newcommand{\p}[1]{(\ref{#1})}
\newcommand{\R}{\mbox{\rule{0.2mm}{2.8mm}\hspace{-1.5mm} R}}
\newcommand{\Z}{Z\hspace{-2mm}Z}
\newcommand{\cd}{{\cal D}}
\newcommand{\cg}{{\cal G}}
\newcommand{\ck}{{\cal K}}
\newcommand{\cw}{{\cal W}}
\newcommand{\vj}{\vec{J}}
\newcommand{\vl}{\vec{\lambda}}
\newcommand{\vz}{\vec{\sigma}}
\newcommand{\vt}{\vec{\tau}}
\newcommand{\poiss}{\stackrel{\otimes}{,}}
\newcommand{\tx}{\theta_{12}}
\newcommand{\tb}{\overline{\theta}_{12}}
\newcommand{\zw}{{1\over z_{12}}}
\newcommand{\sqp}{{(1 + i\sqrt{3})\over 2}}
\newcommand{\sqm}{{(1 - i\sqrt{3})\over 2}}
\newcommand{\NP}[1]{Nucl.\ Phys.\ {\bf #1}}
\newcommand{\PLB}[1]{Phys.\ Lett.\ {B \bf #1}}
\newcommand{\PLA}[1]{Phys.\ Lett.\ {A \bf #1}}
\newcommand{\NC}[1]{Nuovo Cimento {\bf #1}}
\newcommand{\CMP}[1]{Commun.\ Math.\ Phys.\ {\bf #1}}
\newcommand{\PR}[1]{Phys.\ Rev.\ {\bf #1}}
\newcommand{\PRL}[1]{Phys.\ Rev.\ Lett.\ {\bf #1}}
\newcommand{\MPL}[1]{Mod.\ Phys.\ Lett.\ {\bf #1}}
\newcommand{\BLMS}[1]{Bull.\ London Math.\ Soc.\ {\bf #1}}
\newcommand{\IJMP}[1]{Int.\ J.\ Mod.\ Phys.\ {\bf #1}}
\newcommand{\JMP}[1]{Jour.\ Math.\ Phys.\ {\bf #1}}
\newcommand{\LMP}[1]{Lett.\ Math.\ Phys.\ {\bf #1}}
\renewcommand{\thefootnote}{\fnsymbol{footnote}}
\newpage
\setcounter{page}{0}
\pagestyle{empty}
\vs{12}
\begin{center}
{\LARGE {\bf Triality of Majorana-Weyl Spacetimes}}\\
{\LARGE {\bf with Different Signatures}}\\
[0.8cm]

\vs{10} {\large F.
Toppan} ~\\ \quad \\
 {\em CBPF, DCP, Rua Dr. Xavier Sigaud
150, cep 22290-180 Rio de Janeiro (RJ), Brazil}\\
\end{center}
\vs{6} \centerline{\em Talk given at the Workshop on Super and
Quantum Symmetries.}
\centerline{\em Dubna, July 1999.} \vs{6}

\centerline{ {\bf Abstract}}

\vs{6}

Higher dimensional Majorana-Weyl spacetimes present space-time
dualities which are induced by the $Spin(8)$ triality
automorphisms. This corresponds to a very fundamental property of
the supersymmetry in higher dimensions, i.e. that any given theory
can be formulated in different signatures all interconnected by
the $S_3$ permutation group. \vs{6} \vfill
\rightline{CBPF-NF-063/99} {\em E-mail: toppan@cbpf.br}

\section{Introduction.}

Physical theories formulated in different-than-usual spacetimes
signatures have recently found increased attention. One of the
reasons can be traced to the conjectured $F$-theory \cite{Vafa}
which supposedly lives in $(2+10)$ dimensions \cite{Nish}.
The current interest in AdS theories motivated by the AdS/CFT
correspondence furnishes another motivation. Two-time physics
e.g. has started been explored by Bars and collaborators in a
series of papers \cite{Bars}. For another motivation we can also
recall that a fundamental theory is
expected to explain not only the spacetime dimensionality, but
even its signature (see \cite{Duff}). Quite recently Hull and
Hull-Khuri \cite{Hull} pointed out the existence of dualities
relating different compactifications of theories formulated in
different signatures. Such a result provides new insights to the
whole question of spacetime signatures.
Other papers (e.g.\cite{Cori}) have remarked the
existence of space-time dualities.\par
Majorana-Weyl spacetimes (i.e. those supporting Majorana-Weyl
spinors) are at the very core of the present knowledge of
the unification via supersymmetry, being at the basis of
ten-dimensional superstrings, superYang-Mills and
supergravity theories (and perhaps the already mentioned
$F$-theory). A well-established feature of Majorana-Weyl
spacetimes is that they are endorsed of a rich structure.
A legitimate question that could be asked is whether they are
affected, and how, by space-time dualities. The answer can be
stated as follows, all different Majorana-Weyl spacetimes
which are possibly
present in any given dimension are each-other related by duality
transformations which are induced by the $Spin(8)$ triality
automorphisms.
The action of the triality automorphisms is quite non-trivial and
has far richer consequences than the ${\bf Z}_2$-duality
(its most trivial representative) associated to the space-time
$(s,t)\leftrightarrow (t,s)$ exchange discussed in \cite{Duff}.
It corresponds to $S_3$, the six-element group of permutations
of three letters, identified with the group of congruences of the
triangle and generated by two reflections.
The lowest-dimension in which the triality action is non-trivial
is $8$ (not quite a coincidence), where the spacetimes
$(8+0)-(4+4)-(0+8)$ are all interrelated. They correspond to the
transverse coordinates of the $(9+1)-(5+5)-(1+9)$ spacetimes
respectively, where the triality action can also be lifted. Triality
relates as well the $12$-dimensional Majorana-Weyl spacetimes
$(10+2)-(6+6)-(2+10)$, i.e. the potentially interesting cases for
the $F$-theory. Triality allows explaining the presence of points
(read theories) in the brane-molecule table of ref. \cite{Duff},
corresponding to the different versions of e.g. superstrings,
$11$-dimensional supermembranes, fivebranes.\\
As a consequence of triality, supersymmetric theories formulated
with Majora-na-Weyl spinors in a given dimension but with different
signatures, are all dually mapped one into another.
A three-language dictionary can be furnished with the exact
translations among, e.g., the different versions of
such supersymmetric theories.\par
It should be stressed the fact that, unlike \cite{Hull}, the dualities
here discussed are already present for the {\it uncompactified}
theories and in this respect look more fundamental
The reason why the triality of the $d=8$-dimension plays a role
even for Majorana-Weyl spacetimes in $d>8$
is in consequence of the representation properties of
$\Gamma$-matrices in higher dimensions.

\vspace{0.2cm} \noindent{\section{The set of data for
Majorana-Weyl supersymmetric theories.}

At first we present the set of data needed to specify a
supersymmetric theory involving Majorana-Weyl spinors.
The notations here introduced follow \cite{Kugo}
and \cite{DeAn}.\par
 The most
suitable basis is the Majorana-Weyl basis (MWR),
where all spinors are either real or
imaginary. In such a representation the following set of data
underlines any given theory:\\ {\em{\bf i)}} the vector fields
(or, in the string/brane picture, the bosonic coordinates of the
target $x_m$), specified by a vector index denoted by $m$;\\
{\em{\bf ii)}} the spinor fields (or, in the string/brane picture,
the fermionic coordinates of the target $\psi_a$,
$\chi_{\dot{a}}$), specified by chiral and antichiral indices $a$,
$\dot{a}$ respectively;\\ {\em{\bf iii)}} the diagonal
(pseudo-)orthogonal spacetime metric $(g^{-1})^{mn}$, $g_{mn}$
which we will assume to be flat;\\ {\em{\bf iv)}} the ${\cal A}$
matrix, used to define barred spinors,
coinciding with the $\Gamma^0$-matrix in the Minkowski case; in a
Majorana-Weyl basis is decomposed in an equal-size block diagonal
form such as
${\cal A} = A\oplus {\tilde A}$, with structure of indices
${(A)_a}^b$ and ${(\tilde{A})_{\dot{a}}}^{\dot{b}}$
respectively;\\ {\em{\bf v)}} the charge-conjugation matrix ${\cal
C}$ which also appears in an equal-size block diagonal form ${\cal
C} = {C^{-1}}\oplus {\tilde C}^{-1}$. It is invariant under
bispinorial transformations and it can be promoted to be a metric
in the space of chiral (and respectively antichiral) spinors, used
to raise and lower spinorial indices. Indeed we can set
$(C^{-1})^{ab}$, $(C)_{ab}$, and $({\tilde
C}^{-1})^{\dot{a}\dot{b}}$, $({\tilde C})_{\dot{a}\dot{b}}$;\\
{\em {\bf vi)} } the $\Gamma$-matrices, which are decomposed in
equal-size blocks, $\sigma^m$'s the
upper-right blocks and ${\tilde \sigma}^m$'s the lower-left blocks
having structure of indices ${(\sigma^m)_a}^{\dot{b}}$ and
${(\tilde{\sigma}^m)_{\dot{a}}}^b$ respectively;\\ {\em{\bf vii)}}
the $\eta=\pm 1$ sign, labeling the two inequivalent choices for
${\cal C}$.\par The above structures are common
in any theory involving Majorana-Weyl spinors. An explicit
dictionary relating Majorana-Weyl spacetimes having the same
dimensionality but different signature is presented in \cite{DRT}.
The
structures {\em\bf i)}}-{\em{\bf vii)}} are related via
triality transformations which close the $S_3$ permutation group.
They constitute the ``words" in the three-language dictionary.
\par
Majorana-Weyl spacetimes exist for different
signatures of a given dimension $d$ if $d\geq 8$. The special $d=8$
case is the fundamental one. Indeed, we are able to express
higher-dimensional $\Gamma$ matrices
(and in consequence all the above-mentioned structures which define a
Majorana-Weyl theory)
in terms of lower-dimensional ones
according to the recursive formula
\begin{eqnarray}
{\Gamma_d}^{i=1,..., s+1} &=& \sigma_x\otimes {\bf 1}_L\otimes
{\Gamma_s}^{i=1,...,s+1}\nonumber\\
{\Gamma_d}^{s+1+j = s+2, ..., d} &=&
\sigma_y\otimes{\Gamma_r}^{j=1,...,r+1}\otimes {\bf 1}_R
\label{algo}
\end{eqnarray}
where ${\bf 1}_{L,R}$ are the unit-matrices in the respective
spaces, while $\sigma_x = e_{12}+e_{21}$ and $\sigma_y = -i e_{12}+i
e_{21}$ are the $2$-dimensional Pauli matrices.
${\Gamma_r}^{r+1}$ corresponds to the ``generalized $\Gamma^5$-matrix"
in $r+1$ dimensions. In the above
formula the values $r,s=0$ are allowed. The corresponding
${\Gamma_0}^1$ is just $1$.\par
With the help of this formula we are able to reduce the analysis
of different-signatures Majorana-Weyl spacetimes to the $8$-dimensional
case.
In this particular dimension the three
indices, vector ($m$), chiral ($a$) and antichiral ($\dot a$) take
values $m,a , {\dot a} \in \{ 1,...,8\}$.\par
The three Majorana-Weyl solutions, for signatures $(4+4)$,
$(8+0)$, $(0+8)$ find a representation in a Majorana-Weyl basis
with definite (anti-)symmetry property of the $\Gamma$ matrices. In
particular for
the $(4+4)$-signature the $(4_S+4_A)$-representation (see \cite{DRT})
of the
$\Gamma$-matrices has to be employed for both values of $\eta$ in
order to provide a Majorana-Weyl basis. In the ($t=8$, $s=0$)
signature the $(8_S+0_A)$-representation offers a MW basis for
$\eta =+1$, while the $(0_S+8_A)$ offers it for $\eta = -1$. The
converse is true in the ($t=0$, $s=8$)-signature.\par

\vspace{0.2cm}
\noindent{\section{Trialities.}}

The $S_3$ outer automorphisms of the $D_4$ Lie algebra is responsible
for the triality property among the $8$-dimensional vector, chiral
and antichiral spinor representations of $SO(8)$ and $SO(4,4)$
which has been first discussed by Cartan \cite{Cart}.
However, besides such Cartan's V-C-A triality, other triality
related properties follow as a consequence.
For purpose of clarity it will be convenient to
represent them symbolically with triangle diagrams.\par
A first extra-consequence of triality appears
at the level of Majorana-Weyl type of representations for Clifford's
$\Gamma$-matrices.
Such representations can be defined as those where all $\Gamma$'s matrices
exhibit a well-defined
(anti-)symmetry property. In dimension $d=8$ three different
representations of this kind exist (they have been mentioned
in the previous section).
Such different
eight-dimensional representations can be accomodated into
the triangle diagram
\begin{eqnarray}
&\begin{array}{ccccc}
   &  & (4_S+4_A) &  &\\
   & \circ&  &\circ &\\
  (8_S+0_A) &  &\circ&  &(0_S+8_A)\\
\end{array}&
\end{eqnarray}
exhibiting the triality operating at the level
of $\Gamma$-matrices. The $S_3$ transformations relating the three
above representations are realized by similarity transformations. They depend
on the concrete choice of the $\Gamma$-matrix representatives and
will not been furnished here (see however \cite{DRT}).\par
We have already recalled that such
MW-representations are associated with the space-time signature,
due to the fact that the introduction of a Majorana-Weyl basis for
spinors requires the use of the corresponding Majorana-Weyl
representation for Clifford's $\Gamma$ matrices.
As a consequence the triality
can be lifted to operate on the whole set of data introduced in the
previous section; it can therefore be regarded as
operating on the different space-times signatures which support
Majorana-Weyl spinors in a given dimensionality, according to
the triangles
\begin{eqnarray}
\left(\begin{array}{ccccc}
   &  & 5+5 &  &\\
   & \circ&  &\circ &\\
  9+1 &  &\circ&  &1+9\\
\end{array}\right)
&\mapsto &\left(\begin{array}{ccccc}
   &  & 4+4 &  &\\
   & \circ&  &\circ &\\
  8+0 &  &\circ&  &0+8\\
\end{array}\right)\nonumber\\
\end{eqnarray}
The arrow has been inserted to recall that such triality can be
lifted to higher dimensions or, conversely, that the
$8$-dimensional spacetimes arise as transverse coordinates spaces
in some physical theories (the most natural example is the $10$-dimensional
superstring).\par
The Cartan's V-C-A triality, schematically represented as
\begin{eqnarray}
\left(\begin{array}{ccccc}
   &  & V &  &\\
   & \circ&  &\circ &\\
  C &  &\circ&  & A\\
\end{array}\right)
\end{eqnarray}
and the signature triality can also be combined and symbolically
represented by a sort of fractal-like double-triality diagram as
follows
\begin{eqnarray}
&\begin{array}{ccccc}
   &  &
\begin{array}{ccccc}
   &  & V &  &\\
   & \cdot&{\large{\bf 4+4}}  &\cdot &\\
  C &  &\cdot&  &A\\
\end{array}
&  &\\ & &  & &\\ & &  & &\\ &\circ &  & \circ&\\ & &  & &\\ & &
& &\\
\begin{array}{ccccc}
&  & V &  &\\ & \cdot&{\large{\bf 8+0}}  &\cdot &\\ C &  &\cdot&
&A\\
\end{array}
&  &\circ &  &
\begin{array}{ccccc}
&  & V &  &\\& \cdot&{\large{\bf 0+8}}  &\cdot &\\ C &  &\cdot&
&A\\
\end{array}
\\
\end{array}&
\nonumber\\&&\nonumber\\&&
\end{eqnarray}
The bigger triangle illustrates the signature triality, while the
smaller triangles visualize the trialities for vectors, chiral and
antichiral spinors which can be accomodated in each space-time.
\par
It is worth stressing the fact that
the arising of the $S_3$ permutation group as a signature-duality
group for Majorana-Weyl spacetimes in a given dimension {\em is
not} a completely straightforward consequence of the existence of
Majorana-Weyl spacetimes in three different signatures. Some
extra-requirements have to be fulfilled in order to reach this
result. As an example we just mention that a necessary condition
for the presence of $S_3$ requires that each given couple of the
three different spacetimes must differ by an {\em even} number of
signatures (in \cite{DRT} this point is discussed in full
detail); the flipping of an {\em odd} number of signatures, like
the Wick rotation from Minkowski to the Euclidean space, cannot
be achieved with a ${\bf Z}_2$ group when spinors are involved.
An example is provided by the fact
that the change of signature e.g. from $(++)\mapsto (--)$ can be
realized on $\Gamma$-matrices through similarity trasformations
expressed in terms of the
$\sigma_y$ Pauli
matrix $\sigma_y = -i e_{12} +i e_{21}$, through
\begin{eqnarray}
\sigma_y \cdot{\bf 1}_2 \cdot {\sigma_y}^T &=& - {\bf 1}_2
\end{eqnarray}
Of course $\sigma_y$ satisfies ${\sigma_y}^2= {\bf 1}_2$
and therefore it closes a ${\bf Z}_2$ group. On the contrary,
a standard Wick rotations from the Minkowski to the Euclidean
space leads to a ${\bf Z}_4$ group
when represented on $\Gamma$ matrices.\par
Similarity transformations realized by $\sigma_y$ Pauli matrices are
among the building blocks for constructing the $S_3$ duality
transformations for different signature Majorana-Weyl spacetimes.
The formulas will not be reproduced here (they are furnished in
\cite{DRT}, together with the details of the construction).\par
Let us conclude this section by mentioning that triality can be seen not
only as a source of duality-mappings, but as an invariance property. In
the original Cartan's formulation \cite{Cart} this is seen as follows.
At first a group ${\cal G}$ of invariance is introduced as the group of
linear homogeneous transformations acting on the $8\times 3=24$ dimensional
space leaving invariant, separately, the  bilinears ${\cal B}_V$,
${\cal B}_C$, ${\cal B}_A$ for vectors, chiral and antichiral spinors
respctively
(the spinors are assumed commuting in this case) plus a trilinear term
${\cal T}$. Next, the triality group ${\cal G}_{Tr}$ is defined by
relaxing one condition, as the group of linear homogeneous transformations
leaving invariant ${\cal T}$ and the total bilinear ${\cal B}_{Sum}$,
\begin{eqnarray}
{\cal B}_{Sum} &=& {\cal B}_V + {\cal B}_C +{\cal B}_A
\end{eqnarray}\\
It can be proven that ${\cal G}_{Tr}$ is given by the semidirect product of
${\cal G}$ and $S_3$:
\begin{eqnarray}
{\cal G}_{Tr} &=& {\cal G} \otimes_S S_3
\nonumber
\end{eqnarray}
This feature can be extended to the other aspects of triality. It follows
the possibility to look at formulations of higher dimensional
supersymmetric theories presenting an $S_3$ group of symmetry under the
exchange of space and time coordinates.
\par
It should be mentioned that the higher-dimensional supersymmetry strongly
restricts the class of finite groups which can provide ``unification
between space and time" or, otherwise stated, symmetry under time-versus-space
coordinates exchange. In the bosonic case such class of groups is quite large,
while if we consider e.g. the $10$-dimensional supersymmetric case only three
possibilities are left, namely
{\bf i)} the identity ${\bf 1}$, corresponding to a theory formulated in the
single spacetime $(5,5)$,
{\bf ii)} the ${\bf Z}_2$ group for a theory which is formulated by using
two spacetimes copies $(1,9)$ and $(9,1)$,
{\bf iii)} the $S_3$ group; whose corresponding ``space-time unified" theory
requires the introduction of the whole set of three $10$-dimensional
Majorana-Weyl spacetimes $(1,9)$, $(5,5)$, $(9,1)$.

\vspace{0.2cm}
\noindent{\section{Conclusions.}}

In this paper we have shown that the triality automorphisms of
$Spin(8)$, besides its consequences on the representation properties
of the $8$-dimensional vectors, chiral and antichiral spinors (the
usual Cartan's notion of triality), can be realized on
classes of $\Gamma$-matrices representations which furnish a
Majorana-Weyl basis for Majorana-Weyl spinors. Next, triality
transformations can be lifted to connect spacetimes supporting
Majorana-Weyl spinors sharing the same dimensionality, but different
signatures. Recursive formulas for $\Gamma$-matrix representations allow
to extend the $8$-dimensional properties to higher-dimensional cases
as well. Dualities induced by triality are found connecting even-dimensional
Majorana-Weyl spacetimes (and odd-dimensional Majorana ones).
The presence of different formulations of e.g. brane theories, as shown in
the brane-scan molecule table of ref. \cite{Duff} arises as a consequence.\par
Indeed higher dimensional supersymmetric theories admits formulations in
different signatures which are all interrelated by triality induced
transformations.\par
Besides this action of triality as a source of duality mappings between
different versions of supersymmetric theories, triality can provide a setting
to discuss formulation of theories invariant under space-versus-time
coordinates exchange. This would amount to investigate the formulation of
supersymmetric theories exhibiting a manifest $S_3$-invariance under
signature-triality transformations.
\par
The range of possible applications for the methods and the ideas here
discussed is vast. Let us just mention that are currently
investigated the web of dualities connecting the six different
versions of the $12$-dimensional
Majorana-Weyl spacetimes which should support the $F$-theory
(the number $6=3\times 2$ is due to the three different signatures
of Majorana-Weyl spacetimes and the two values of the $\eta$ sign), with the
$6$ versions of the $11$-dimensional Majorana spacetimes (for the
$M$-theory) in $(10+1)$, $(9+2)$, $(6+5)$, $(5+6)$, $(2+9)$,
$(1+10)$ signatures and with the different (again $3\times 2$)
versions of the $10$-dimensional Majorana-Weyl spacetimes.

\vskip0.6cm \noindent{\Large{\bf Acknowledgments}} \\ {\quad}\\
The talk here presented is mainly based on a work
with M. A. De Andrade and M. Rojas, who I am very pleased
to acknowledge for the fruitful collaboration.

\end{document}